Libertas Academica
FREEDOM TO RESEARCH

REVIEW

# Some Perspectives on Network Modeling in Therapeutic Target Prediction

Reka Albert[1], Bhaskar DasGupta[2] and Nasim Mobasheri[2]

[1]Department of Physics, Pennsylvania State University, University Park, PA. [2]Department of Computer Science, University of Illinois at Chicago, Chicago, IL. Corresponding author email: ralbert@phys.psu.edu, bdasgup@uic.edu

**Abstract:** Drug target identification is of significant commercial interest to pharmaceutical companies, and there is a vast amount of research done related to the topic of therapeutic target identification. Interdisciplinary research in this area involves both the biological network community and the graph algorithms community. Key steps of a typical therapeutic target identification problem include synthesizing or inferring the complex network of interactions relevant to the disease, connecting this network to the disease-specific behavior, and predicting which components are key mediators of the behavior. All of these steps involve graph theoretical or graph algorithmic aspects. In this perspective, we provide modelling and algorithmic perspectives for therapeutic target identification and highlight a number of algorithmic advances, which have gotten relatively little attention so far, with the hope of strengthening the ties between these two research communities.

**Keywords:** drug target identification, complex network of interactions, Boolean models, signal transduction networks, transitive reduction, node essentiality

This article is available from http://www.la-press.com.

## Introduction

We begin with a motivating example to illustrate a typical therapeutic target design issue that is of interest to this paper. The blood cancer T-cell large granular lymphocyte (T-LGL) leukemia exhibits an abnormal proliferation of mature cytotoxic T lymphocytes (CTLs). CTLs are generated to eliminate cells infected by a virus, but unlike normal CTLs, which undergo activation-induced cell death after they successfully fight the virus, leukemic T-LGL cells remain long-term competent.[1] To date, 33 proteins and small molecules related to cytotoxic T lymphocyte activation and activation-induced cell death have been shown to be deregulated (eg, constitutively active) in T-LGL. It is known that prosurvival signalling pathways such as the MAPK and JAK-STAT3 pathways are upregulated and that T-LGL cells are insensitive to Fas-induced apoptosis (programmed cell death).[2] As there is no known curative therapy for T-LGL leukemia, identification of a potential therapeutic target, for example, a protein whose knockdown would lead to apoptosis of these T-LGL cells, is of high priority.

The above example highlights some key steps of our typical therapeutic target identification problem, namely, we need to synthesize or infer the complex network of interactions relevant to the disease (in this case, the signalling network corresponding to activation-induced cell death), we need to connect this network to the relevant behavior (in this case, aberrant survival), and we need to predict which components are key mediators of the behavior. All of these steps involve nontrivial graph algorithmic aspects. In this review, we aim to highlight a number of algorithmic advances that have gotten relatively little attention so far.

## Models of Interactions

We assume that the reader is familiar with basic graph algorithmic concepts[3] as well as standard biological terminology in bioinformatics.[4] We begin by recalling two frequently used network representations for interacting biological systems.

## Boolean model

The Boolean model is typically used in the context of studying dynamics of biological networks in which the measurement of quantities of interest (eg, gene expression levels) is binarized (eg, the expression level is either 1 or 0 indicating if the gene is expressed or not) or relative (eg, the expression level in condition one is higher than that in condition two). The state and input variables are Boolean variables, and the function that updates the state of a variable is also a Boolean function (of other variables). Figure 1 provides a pictorial illustration of a Boolean model.

## Regulatory network model

Formally, we define a regulatory network to be an edge-labelled directed graph in which nodes represent individual components of the biological system and directed edges of the form $(u,v)$ indicates that node $u$ has an influence on node $v$. The edge label $\ell(u,v)$ of an edge $(u,v)$ indicates the nature of the causal relationship, with $\ell(u,v) = 1$ and $\ell(u,v) = -1$ indicating that $u$ has an excitatory (positive) and inhibitory (negative) influence on $v$, respectively (see Fig. 2). This representation applies to gene regulatory networks (describing the regulation of gene transcription and related processes) and signal transduction networks (describing the information flow from external signals to within-cell components).

## Synthesizing Signal Transduction Networks

Signal transduction and gene regulatory networks are crucial to the maintenance of cellular homeostasis and for cell behavior such as growth, survival, apoptosis, and movement. Deregulation of these networks is a key contributor to many disease processes such as developmental disorders,[5] diabetes,[6] vascular diseases,[5,7] autoimmunity,[8] and cancer.[5,9,10] For example, gene mutations or expression changes can lead to incorrect behaviors that result in tumour development and/or the promotion of cell migration and metastasis.[11] Thus identification of the regulatory

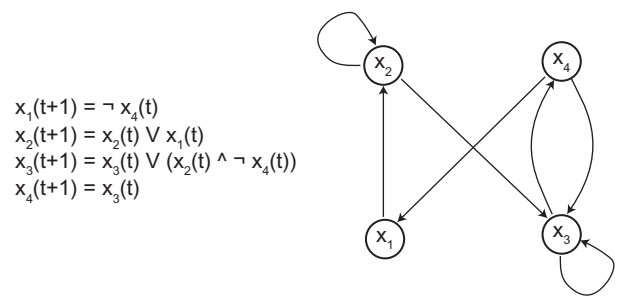

$x_1(t+1) = \neg \, x_4(t)$
$x_2(t+1) = x_2(t) \lor x_1(t)$
$x_3(t+1) = x_3(t) \lor (x_2(t) \land \neg \, x_4(t))$
$x_4(t+1) = x_3(t)$

**Figure 1.** A Boolean model with 4 state variables.

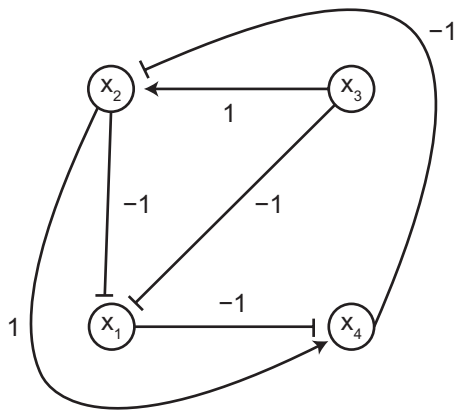

**Figure 2.** A signal transduction network.

networks responsible for cellular behaviors relevant to a certain disease is a crucial first step toward the development of therapeutic strategies targeting that disease. A key feature of this network is that it should contain a node (or a small set of nodes) that can be designated as a proxy or proxies for the disease. This node may be a concept node describing a behavior such as apoptosis or migration, and its upstream nodes (regulators) should be the proteins most directly responsible for the behavior, for example, caspases involved in programmed cell death. In the case of the T-LGL network, wherein apoptosis is the normal outcome, the best proxy for the disease should be a node that embodies the logical negation of apoptosis. Its upstream regulators are the same nodes as the upstream regulators of the node apoptosis, but the sign (label) of the edges is reversed.

The synthesis of such a network starts with extensively collecting the relevant literature and experimental data concerning the interactions among the elements of the network. Although for many biological processes different experiments generate an abundance of relevant components and causal interactions, there is insufficient information on the overall structure and mechanisms of these processes. Therefore, information sources from individual experiments need to be assembled and integrated.

Experimental evidence about the involvement of a component or a regulatory relationship in a biological process has several types. For example, a concentration change of a protein after treating the system with an input signal indicates that this protein may be a component of this signal's signal transduction network. Such evidences can be collected from high-throughput gene expression, proteomics, and metabolomics data. In addition, if knocking out or overexpressing a component leads to a change in the relevant cellular response, it can be concluded that this component is involved in the biological process. Causal relationships between components can be collected from high-throughput phospho-proteomics, protein-DNA interaction, as well as genetic interaction studies. These causal relationships can be represented as directed edges from one component to another characterized by one of two signs: activating (positive) or inhibitory (negative). However, in some cases, perturbation experiments lead to composite causal relationships, which need to be broken down to component-to-component relationships depending on the concrete situation. In all but a handful of well-studied cases, a very useful step in network analysis is to find the most parsimonious (minimal) network that explains all the experimentally obtained causal observations.

The generation of this most parsimonious network from an initially too-redundant network involves systematically removing those edges or nodes that do not change "appropriate" reachability relations between nodes. For this purpose, we start by describing two necessary network optimization problems.

## Binary transitive reduction and pseudo node collapse

For these two problems, the notation $u \Rightarrow_x v$ is used to denote a (directed) path $P$ from $u$ to $v$ of parity $x = \Pi_{e \in P} \ell(e) \in \{-1, 1\}$ (an edge is simply denoted by $u \rightarrow_x v$).

The first optimization problem is the binary transitive reduction (BTR) problem.[12–14] In this problem, our goal is to find a minimal subgraph from the original nonminimal signal transduction network by removing the "redundant" edges, namely those edges $u \rightarrow_x v$ for which an alternate pathway $u \Rightarrow_x v$ not using the edge $u \rightarrow_x v$ exists. See Figure 3 for an illustration. The goal of BTR is to produce a network topology that is as close as possible to a tree topology while supporting all experimental observations. The implicit assumption of tree-like topologies permeates the traditional molecular biology literature: signal transduction and metabolic pathways were assumed to be close to linear chains, and genes were assumed to be regulated by one or two transcription factors.[15]

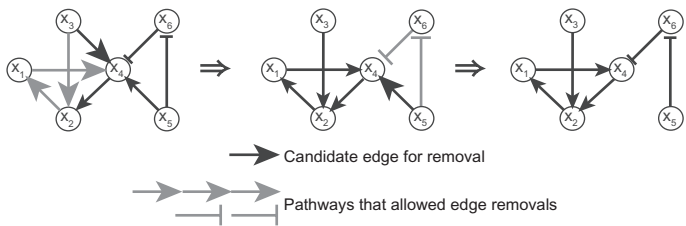

→ Candidate edge for removal

⇢ ⊣ Pathways that allowed edge removals

**Figure 3.** An illustration of BTR.

According to current observations, the reality is not far: the average in/out degree of transcriptional regulatory networks[16,17] and the mammalian signal transduction network[18] is close to 1.

The second optimization problem is the (generalized) pseudo-node-collapse (PNC) problem.[13,14] An instance of PNC includes a given subset $V_{pseudo} \subseteq V$ of nodes called pseudo nodes. For any node $v$, define $in(v) = \{(u,x) \mid u \Rightarrow_x v\}\setminus\{v\}$ and $out(v) = \{(u,x) \mid v \Rightarrow_x u\}\setminus\{v\}$. Collapsing 2 nodes $u$ and $v$ is permissible provided at least one of them is a pseudo node, $in(u) = in(v)$ and $out(u) = out(v)$. If permissible, the collapse of 2 nodes $u$ and $v$ creates a new vertex $w$, makes every incoming (respectively, outgoing) edge to (respectively from) either $u$ or $v$ an incoming (respectively, outgoing) edge from $w$, removes any parallel edge that may result from the collapse operation, and also removes both nodes $u$ and $v$. Our goal is to obtain a minimal graph $G' = (V',E')$ from $G$ by a sequence of permissible collapse operations such that $G'$ is consistent, that is, for no pairs of nodes $u$ and $v$ both $u \Rightarrow_1 v$ and $u \Rightarrow_{-1} v$ exists.

## Network construction

Following the approaches in Kachalo et al,[13] Albert et al,[14] and Albert et al,[19] interaction information between components can be partitioned into two main categories: (1) *Direct* interactions related to biochemical evidences that provide information on enzymatic activity or protein-protein interactions. (2) Putative interaction patterns (double-causal relationships) that arise from the observation of differential responses following perturbation experiments. For example, a decreased response (R) to a stimulus (S) in an organism wherein component X was knocked out, as compared to the normal (wild-type) organism, leads to the double-causal relationship $X \Rightarrow_x (S \Rightarrow_y R)$.

Kachalo et al,[13] Albert et al,[14] and Albert et al[19] describe a general methodology for synthesizing direct and double-causal information into a minimal consistent network. The methodology is shown in Figure 4. A software named NET-SYNTHESIS for synthesis of networks based on the framework in Figure 4 was reported in Kachalo et al[13] and Albert et al;[19] this software was used to build a network model to study the signalling components that effect the survival of cytoxic T lymphocytes in LGL Leukemia.[20]

## Pathway preserving network simplification

PNC can also be used in a much broader context of network simplification in the following manner. In many large-scale regulatory networks, only a subset of the nodes are of inherent interest (eg, because they are differentially expressed in different exogenous conditions), and the rest serve as background or mediators. One can therefore designate nodes of less interest or confidence as pseudo nodes and then collapse them, thereby making the network among high-interest/confidence nodes easier to interpret. For example, using this approach, PNC with BTR can be used to focus on specific pathways in disease networks to better understand the molecular mechanism of the onset of the disease and therefore help in drug target identification.[13]

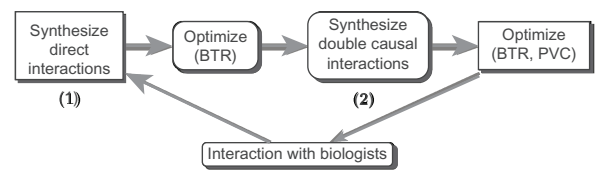

**(1)** Synthesize direct interactions → Optimize (BTR) → **(2)** Synthesize double causal interactions → Optimize (BTR, PVC)

Interaction with biologists

**(1) [Encoding direct interactions]**

Biochemical and pharmacological evidences that define component-to-component relationships, namely of the form "A promotes B" or "A inhibits B", are incorporated (in arbitrary order) as directed edges $A \rightarrow_y B$ or $A \rightarrow_{-y} B$, respectively. If the interaction is known to be a direct interaction with concrete evidence, then the edge is marked as a *critical edge* and included in $E_c$.

**(2) [Encoding double causal interactions]**

Consider each double causal interaction $A \rightarrow_x (B \rightarrow_y C)$, where $x,y \in \{-1, 1\}$, in an arbitrary order. Add new nodes and/or edges in the network based on the following cases:

• If $B \rightarrow_y C \in E_c$, then add the edge $A \rightarrow_x B$.

• Otherwise, if there is no subgraph (in the network constructed so far) of the form

$$A \\ \Downarrow_x \quad \text{for some node } D \text{ where } b = ab = y, \text{ then add}$$
$$B \Rightarrow_x D \Rightarrow_b C$$

the subgraph

$$A \\ \downarrow_x \quad \text{to the network where } P \text{ is a new pseudo-}$$
$$B \Rightarrow_x P \rightarrow_b C$$

node and $b = a b = y$.

**Figure 4.** A framework for network synthesis using BTR and PNC.[16–18,23]

As an illustration, this approach was used by Kachalo et al[13] to focus on pathways that involve the 33 known T-LGL deregulated proteins in a cell-survival/cell-death regulation related signalling network synthesized from the Transpath 6.0 database.

The approach used by Kachalo et al was to focus special interest on the effect of Ras on apoptosis response through Fas/FasL pathway by designating nodes that correspond to proteins with no evidence of being changed during this effect as pseudo nodes and simplifying the network via iterations of PNC and BTR. Although performing comprehensive PNC in this manner may lead to a drastic reduction of the network, the drawback of such a dramatic simplification is that pairs of incoherent edges (two edges with opposite labels) can appear among the same pair of nodes. While incoherent paths between pairs of nodes are often seen in biological regulatory networks, interpretation of incoherent edges is difficult without knowledge of the mediators of the two opposite regulatory mechanisms. Thus, optimal simplification in this manner may require careful selection of pseudo nodes and PNC algorithms.

## Therapeutic Target Identification

Many approaches to therapeutic target identification are based on analysis of molecular interaction networks. Certain approaches are based on genome-wide reconstructed networks; others aim to first construct the network most relevant to the disease. A powerful view of the first type is that tumor cell types are attractors of the global gene regulatory network.[21] These attractors are normally inaccessible (ie, normal cell processes will not lead to tumor development), but genetic mutations can distort the attractor basins in such a way that a previously nonaccessible attractor is now accessible. The therapeutic target identification then involves finding the perturbations that make the malignant cells revert to the trajectory that leads to the nonmalignant, more differentiated cell.

Approaches of the second type focus on known disease-associated networks, such as growth factor signaling networks, which are frequently deregulated in cancer. Layek et al[22] constructed a Boolean representation of a signaling network that contains five growth factors as inputs and seven proliferative and anti-apoptotic proteins as outputs. Based on this network, they determined the node faults (eg, stuck-at-zero, stuck-at-one) that change the output nodes' states. They also identified the combinations of existing drugs that would best mitigate each fault.

A powerful way of identifying therapeutic targets is to identify the nodes most essential for mediating the abnormal outcome. For this application, the reconstructed signal transduction network relevant to a disease contains an output node that is a proxy for the disease. In graph theoretical terms, essentiality for mediating the abnormal outcome can be expressed as high centrality in the input-output network. A multitude of node centrality measure exists, and several of them, such as closeness centrality,[23] were fruitfully used to predict key nodes in biological regulatory networks. A deeper, dynamic analysis by Abdi et al[24] adapted methods of circuit fault diagnosis engineering to signaling networks to determine the vulnerability of the network to the dysfunction of each node. The nodes with highest vulnerability value were identified as key nodes. One such prediction was validated experimentally. The method proposed by Wang and Albert[25] bridges graph-based and Boolean analysis by developing an enriched graph representation, which we describe next.

### Network expansion

The starting point is to describe each node $v$ by a disjunctive normal Boolean rule

$$
\begin{aligned}
v = \Big( u_{1,1} \ \wedge \ \cdots \ \wedge \ u_{1,n_1} \Big) \\
\vee \ \Big( u_{2,1} \ \wedge \ \cdots \ \wedge \ u_{2,n_2} \Big) \\
\vee \cdots \ \vee \ \Big( u_{m,1} \ \wedge \ \cdots \ \wedge \ u_{m,n_m} \Big)
\end{aligned}
$$

where the $u_{i,j}$ s are regulators of node $v$. The operators connecting the regulators are decided from the literature. In the absence of evidence for conditional dependence (the necessity of two regulators to work together to be effective), the default representation of multiple activating edges converging on the same node is an OR ($\oplus$) relationship. Information on conditional dependence is incorporated by AND ($\otimes$) relationships among edges. Inhibitory regulations are represented by the logical operator NOT. If an inhibitory regulation is dominant among multiple interactions, as is often the case, AND NOT should be used; otherwise one can use OR NOT, which means that the absence of an inhibitor is similar to the presence of an activator.

To incorporate inhibitory interactions, a complementary node is introduced for each node, or at a minimum for each node that negatively regulates other nodes or is being negatively regulated by other nodes. This complementary node represents the logical negation of the original node, and its Boolean rule is thus the negation of the original node's rule. Correspondingly, an inhibitory edge starting at a node becomes an activating edge starting at its complementary node. An inhibitory edge ending at a node becomes an activating edge ending at its complementary node. For each set of conditional interactions (with AND relationship) ending at a node, a composite node is introduced. The regulators of $v$ activate the composite node, which then activates the node $v$.

Introducing complementary nodes and composite nodes increases the number of nodes and edges in the network, but the benefit is that ambiguity is eliminated. All the directed interactions in the expanded network represent activation. All edges ending at a composite node are conditionally dependent on each other, and multiple edges ending in an original node or complementary node are independent.

## Input-output connectivity
Elementary flux modes, that is, minimal sets of enzymes that can make metabolic systems operate at steady states, play an important role in metabolic network analysis.[23] Analogously, an elementary signalling mode (ESM) can be defined as a minimal set of components that can perform signal transduction from initial signals to cellular responses. By minimal, we mean that an ESM is not decomposable and none of its signalling components is redundant, that is, knockout of any of the nodes in the ESM will make it unable to transducer the signal. The concept of ESM is an extension of the graph concept of simple path. An ESM that does not contain any composite nodes is indeed a simple path. If the ESM has a composite node, it additionally includes all the edges ending at the composite node and their source nodes.

ESM identification was implemented in three different ways[25–27]: using an extension of depth-first search (most useful for acyclic networks), an iterative integer linear programming algorithm (most useful for small networks), and a subgraph-growing bottom-up algorithm.

## Node essentiality
The main idea is to rank the importance of signalling components by the effects of their perturbation on the ESMs of the network. It is important to realize that the disruption of any single node may lead to a cascading breakdown of a large part of the system as nodes that lost their indispensable regulators will be lost as well. There are three cases for a regulator $v$ to be indispensable for a direct target node $u$: if $v$ is the sole regulator of $u$, if $u$ is a composite node, or if $v$ is the only remaining regulator of $u$ left due to the disruption of other regulators. The cascading effect of the removal of a node can be determined by an algorithm that iteratively finds and deletes the nodes that have just lost their indispensable regulators. Each node $v$ is then characterized by the relative reduction in the number of ESMs due to the removal of $v$ and its cascading effects.

Wang and Albert[25] apply this method to various signalling networks[20,28–30] and compare the essentiality values with the results of Boolean dynamic models. Components are classified as essential in the dynamic model if their elimination (OFF state) leads to incorrect dynamic behaviors, and nonessential otherwise. Components are classified as essential in the graph method if their importance value is above a threshold. They find high sensitivity (fraction of essential components that are recognized by a method) coupled with high specificity (fraction of nonessential components that are recognized by a method) for a range of thresholds for all the networks. Several specific nodes identified as highly essential are well-supported by experimental observations.[25]

In the specific example of the T-LGL network, 14 components have importance values of one (or very close to one).[31] This means that blocking any of these nodes disrupts (almost) all signalling paths from the complementary node to apoptosis, thus these nodes are candidate therapeutic targets. All of these nodes are also found to be essential for the T-LGL survival state according to a dynamic model, that is, reversing their states causes apoptosis to be the only possible outcome of the system. Moreover, experimental verification of the importance of these nodes exists for 10 of the 14 nodes.[31]

## Conclusion
In this perspective, we have highlighted a few algorithmic advances related to therapeutic identification, which have received relatively little attention so far. We believe that this very important research area will benefit a great deal from interdisciplinary investigations that identify the most fruitful abstractions and develop the computational methodologies most suitable for parsimonious predictive modelling, and we hope that this perspective will strengthen the ties between the biological network and graph algorithm research communities.

## Acknowledgements
The authors thank Eduardo Sontag for useful discussions.

## Author Contributions
Conceived and designed the experiments: RA, BDG, NM. Analysed the data: RA, BDG, NM. Wrote the first draft of the manuscript: RA, BDG, NM. Contributed to the writing of the manuscript: RA, BDG, NM. Agree with manuscript results and conclusions: RA, BDG, NM. Jointly developed the structure and arguments for the paper: RA, BDG, NM. Made critical revisions and approved final version: RA, BDG, NM. All authors reviewed and approved of the final manuscript.

## Funding
B. DasGupta and N. Mobasheri were supported by NSF grants IIS-1160995 and DBI-1062328. R. Albert was supported by NSF grants IIS-1161007 and PHY-1205840.

## Competing Interests
Author(s) disclose no potential conflicts of interest.

## Disclosures and Ethics
As a requirement of publication author(s) have provided to the publisher signed confirmation of compliance with legal and ethical obligations including but not limited to the following: authorship and contributorship, conflicts of interest, privacy and confidentiality and (where applicable) protection of human and animal research subjects. The authors have read and confirmed their agreement with the ICMJE authorship and conflict of interest criteria. The authors have also confirmed that this article is unique and not under consideration or published in any other publication, and that they have permission from rights holders to reproduce any copyrighted material. Any disclosures are made in this section. The external blind peer reviewers report no conflicts of interest. Provenance: the authors were invited to submit this paper.

## References
1. Loughran TP Jr. Clonal diseases of large granular lymphocytes. *Blood*. 1993;82(1):1–14.
2. Epling-Burnette PK, Bai F, Wei S, et al. ERK couples chronic survival of NK cells to constitutively activated ras in lymphoproliferative disease of granular lymphocytes. *Oncogene*. 2004;23:9220–9.
3. Cormen TH, Leiserson CE, Rivest RL, Stein C. *Introduction to Algorithms*. Cambridge, MA: The MIT Press; 2001.
4. Pevzner PA. *Computational Molecular Biology: An Algorithmic Approach*. Cambridge, MA: The MIT Press; 2000.
5. Gordon KJ, Blobe GC. Role of transforming growth factor-β superfamily signaling pathways in human disease. *Biochim Biophys Acta*. 2008;1782: 197–228.
6. Muscogiuri G, Chavez AO, Gastaldelli A, et al. The crosstalk between insulin and renin-angiotensin-aldosterone signaling systems and its effect on glucose metabolism and diabetes prevention. *Curr Vasc Pharmacol*. 2008;6: 301–12.
7. Brasier AR. The nuclear factor-kappaB-interleukin-6 signalling pathway mediating vascular inflammation. *Cardiovasc Res*. 2010;86:211–8.
8. Mavers M, Ruderman EM, Perlman H. Intracellular signal pathways: potential for therapies. *Curr Rheumatol Rep*. 2009;11:378–85.
9. Grzmil M, Hemmings BA. Deregulated signalling networks in human brain tumours. *Biochim Biophys Acta*. 2010;1804:476–83.
10. Ikushima H, Miyazono K. TGFbeta signalling: a complex web in cancer progression. *Nat Rev Cancer*. 2010;10:415–24.
11. Feinberg AP, Ohlsson R, Henikoff S. The epigenetic progenitor origin of human cancer. *Nat Rev Genet*. 2006;7:21–33.
12. Albert R, DasGupta B, Dondi R, Sontag ED. Inferring (biological) signal transduction networks via transitive reductions of directed graphs. *Algorithmica*. 2008;51(2):129–59.
13. Kachalo S, Zhang R, Sontag ED, Albert R, DasGupta B. NET-SYNTHESIS: A software for synthesis, inference and simplification of signal transduction networks. *Bioinformatics*. 2008;24(2):293–5.
14. Albert R, DasGupta B, Dondi R, et al. A novel method for signal transduction network inference from indirect experimental evidence. *J Comput Biol*. 2007;14(7):927–49.
15. Alberts B. *Molecular Biology of the Cell*. New York, NY: Garland Publishers; 1994.
16. Shen-Orr SS, Milo R, Mangan S, Alon U. Network motifs in the transcriptional regulation network of Escherichia coli. *Nat Genet*. 2002;31:64–8.
17. Lee TI, Rinaldi NJ, Robert F, et al. Transcriptional regulatory networks in Saccharomyces cerevisiae. *Science*. 2002;298(5594):799–804.
18. Ma'ayan A, Jenkins SL, Neves S, et al. Formation of Regulatory Patterns During Signal Propagation in a Mammalian Cellular Network. *Science*. 2005;309(5737):1078–83.
19. Albert R, DasGupta B, Sontag ED. Inference of signal transduction networks from double causal evidence. In: Fenyo D, editor. *Methods in Molecular Biology: Topics in Computational Biology*. New York, NY: Springer-Verlag; 2010.
20. Zhang R, Shah MV, Yang J, et al. Network model of survival signaling in LGL leukemia. In: *Proc Natl Acad Sci U S A*. 2008;105(42):239–251.
21. Huang S. On the intrinsic inevitability of cancer: From foetal to fatal attraction. *Semin Cancer Biol*. 2011;21:183–99.
22. Layek R, Datta A, Bittner M, Dougherty ER. Cancer therapy design based on pathway logic. *Bioinformatics*. 2011;27(4):548–55.
23. Latora V, Marchiori M. A measure of centrality based on network efficiency. *New J Phys*. 2007;9:188.
24. Abdi A, Tahoori MB, Emamian ES. Fault diagnosis engineering of digital circuits can identify vulnerable molecules in complex cellular pathways. *Sci Signal*. 2008;1(42):ra10.

25. Wang RS, Albert R. Elementary signaling modes predict the essentiality of signal transduction network components. *BMC Syst Biol.* 2011;5:44.
26. Stelling J, Klamt S, Bettenbrock K, Schuster S, Gilles ED. Metabolic network structure determines key aspects of functionality and regulation. *Nature.* 2002;420:190–3.
27. Wang RS, Sun Z, Albert R. Minimal functional routes in directed graphs with dependent edges. *Int Trans Oper Res.* In press.
28. Thakar J, Pilione M, Kirimanjeswara G, Harvill ET, Albert R. Modeling systems-level regulation of host immune responses. *PLoS Comput Biol.* 2007;3:e109.

29. Li S, Assmann SM, Albert R. Predicting essential components of signal transduction networks: a dynamic model of guard cell abscisic acid signaling. *PLoS Biol.* 2006;4:e312.
30. Saez-Rodriguez J, Simeoni L, Lindquist JA, et al. A logical model provides insights into T cell receptor signaling. *PLoS Comput Biol.* 2007;3:e163.
31. Saadatpour A, Wang RS, Liao A, et al. Dynamical and structural analysis of a T cell survival network identifies novel candidate therapeutic targets for large granular lymphocyte leukemia. *PLoS Comput Biol.* 2011;7:e1002267.